\documentclass[reprint,superscriptaddress,showpacs,epsfig,prl]{revtex4-2}
\usepackage{color,graphicx}
\usepackage{graphics}
\graphicspath{{images/}}
\usepackage{epsfig}
\usepackage{epstopdf}
\usepackage{hyperref}
\usepackage{mathrsfs}
\usepackage{bm}
\usepackage{amsmath,amssymb}
\usepackage{subfiles}
\begin{document}
\title{Fermi surface reconstruction due to the orthorhombic distortion in Dirac semimetal YbMnSb$_2$}
\author{Dilip Bhoi}
\affiliation{The Institute for Solid State Physics, University of Tokyo, Kashiwa, Chiba 277-8581, Japan}
\author{Feng Ye}
\affiliation{Neutron Scattering Division, Oak Ridge National Laboratory, Oak Ridge, Tennessee 37831, USA}
\author{Hanming Ma}
\author{Xiaoling Shen}
\affiliation{The Institute for Solid State Physics, University of Tokyo, Kashiwa, Chiba 277-8581, Japan}
\author{Arvind Maurya}
\affiliation{Department of Physics, School of Physical Sciences, Mizoram University, Aizawl 796 004, India}
\author{Shusuke Kasamatsu}
\affiliation{Academic Assembly (Faculty of Science), Yamagata University, Yamagata 990-8560, Japan}
\author{Takahiro Misawa}
\author{Kazuyoshi Yoshimi}
\affiliation{The Institute for Solid State Physics, University of Tokyo, Kashiwa, Chiba 277-8581, Japan}
\author{Taro Nakajima}
\affiliation{The Institute for Solid State Physics, University of Tokyo, Kashiwa, Chiba 277-8581, Japan}
\author{Masaaki Matsuda}
\affiliation{Neutron Scattering Division, Oak Ridge National Laboratory, Oak Ridge, Tennessee 37831, USA}
\author{Yoshiya Uwatoko}
\affiliation{The Institute for Solid State Physics, University of Tokyo, Kashiwa, Chiba 277-8581, Japan}
\date{\today}
\begin{abstract}
Dirac semi-metal with magnetic atoms as constituents delivers an interesting platform to investigate the interplay of Fermi surface (FS) topology, electron correlation, and magnetism. One such family of semi-metal is YbMn$Pn_2$ ($Pn$ = Sb, Bi), which is being actively studied due to the intertwined spin and charge degrees of freedom. In this Letter, we investigate the relationship between the magnetic/crystal structures and FS topology of YbMnSb$_2$ using single crystal x-ray diffraction, neutron scattering, magnetic susceptibility, magnetotransport measurement and complimentary DFT calculation. Contrary to previous reports, the x-ray and neutron diffraction reveal that YbMnSb$_2$ crystallizes in an orthorhombic $Pnma$ structure with notable anti-phase displacement of the magnetic Mn ions that increases in magnitude upon cooling. First principles DFT calculation reveals a reduced Brillouin zone and more anisotropic FS of YbMnSb$_2$ compared to YbMnBi$_2$ as a result of the orthorhombicity. Moreover, the hole type carrier density drops by two orders of magnitude as YbMnSb$_2$ orders antiferromagnetically indicating band folding in magnetic ordered state. In addition, the Landau level fan diagram yields a non-trivial nature of the SdH quantum oscillation frequency arising from the Dirac-like Fermi pocket. These results imply that YbMnSb$_2$ is an ideal platform to explore the interplay of subtle lattice distortion, magnetic order, and topological transport arising from relativistic quasiparticles.
\end{abstract}
\maketitle
\newpage 
Magnetic Dirac/Weyl semimetals deliver a promising platform, where the novel coupling between magnons and relativistic fermions could be exploited to manipulate the quantum transport phenomena using various parameters like chemical substitution, pressure, strain, etc. In this context, the collinear antiferromagnetic (AFM) ternary $A$Mn$Pn_2$ (where $A$ = rare earth elements like Eu, Yb or alkali earth elements like Ca, Sr, Ba; $Pn$ = pnictides Sb or Bi) have attracted increasing attention due to the presence of anisotropic Dirac cones close to the Fermi level, $E_{\rm{F}}$ \cite{Wang2011, Wang2012, Park2011, Jo2014, Liu2017a, AWang2016, Liu2017b, He2017, Zhang2016, Borisenko2019, Huang2017, Liu2016a, Yi2017a, Fang2014, May2014, Masuda2016}. 

\begin{figure}[th]
	\begin{center}
		\includegraphics[width=0.46\textwidth]{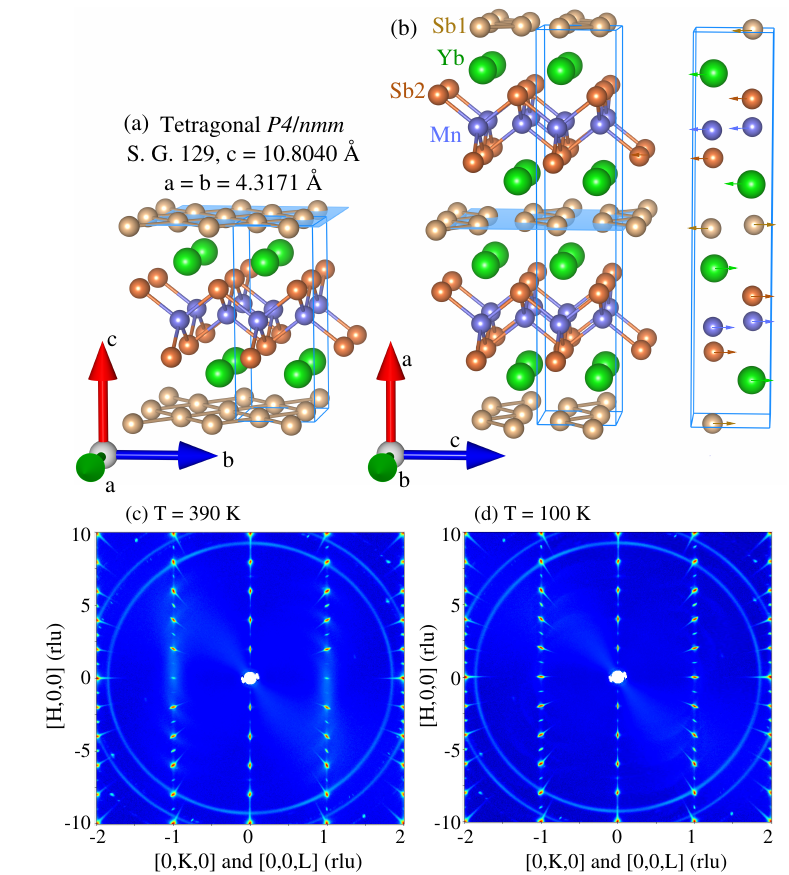}
		\caption{(a) The reported tetragonal $P4/nmm$ crystal structure of YbMnSb$_2$ as in Ref.~\cite{Hu2023,Tobin2023,Soh2021,Qiu2019,Kealhofer2018,Wang2018,Pan2021, Baranets2021}. (b) The orthorhombic $Pnma$ structure reported in this work. Arrows show the displacements of each atom along the $a$-axis of the tetragonal $P4/nmm$ structure. The blue-shaded regions illustrate the Sb$_1$ layers. Blue rectangles highlight the unit cell. The contour plot of the neutron scattering intensity pattern of YbMnSb$_2$ at (c) $T$ = 390 K and (d) $T$ = 100 K in the ($H$,$K$,0) and ($H$,0,$L$) planes using the lattice parameters (21.58 \AA, 4.3 \AA, 4.3 \AA).}\label{SCD}
	\end{center}
\end{figure}

The 112-type $A$Mn$Pn_2$ consists of a stacking of two-dimensional (2D) $Pn$ conduction layers, $A$-layers, and insulating Mn$Pn_4$ layers as shown in Fig.\ref{SCD}(a). In Mn$Pn_4$ layers, each Mn atom is surrounded by four $Pn$ atoms forming a tetrahedron, whereas 2D-$Pn$ layers are responsible mostly for the exotic properties like quantum magnetoresistance \cite{Wang2011, Wang2012, Jo2014, Liu2017a, Park2011,AWang2016} and bulk quantum Hall effect \cite{Masuda2016,Sakai2020,Liu2021}. Band structure calculations revealed that the electronic density of states at $E_{\rm{F}}$ is primarily composed of the $Pn$-$p_{x/y}$ and Mn-$d$ orbitals, suggesting a close relationship between the Mn moment direction and underlying electronic structure \cite{Le2021, Li2019arxiv, Guo2022arxiv, Ni2022, Pan2022}. Interestingly, these calculations further predict that the FM component arising from a canting of Mn moments breaks the time-reversal symmetry, thus playing a vital role in producing different topological states depending on the Mn moment direction \cite{Le2021,Li2019arxiv,Guo2022arxiv, Ni2022,Pan2022,Borisenko2019,Yang2020}. 

Among the 112 materials, YbMn$Pn_2$ are particularly unique due to the coupling between magnetism and Dirac quasiparticles \cite{Saptoka2020, Soh2019, Hu2023, Tobin2023}, unusual interlayer quantum coherent transport \cite{Wang2018, Liu2017a}, including promising attributes required for energy conversion technology like large thermoelectric power \cite{Pan2021, Baranets2021} and giant anomalous Nernst effect \cite{Pan2022}. Band structure calculations \cite{Pan2022,Pan2021,Le2021,Qiu2019,Wang2018,Ni2022} identified YbMn$Pn_2$ as nodal line semimetals, where Fermi surface (FS) consists of two Dirac-like bands and a heavy 3D-parabolic band. Despite these results pointing similar FS topology, experiments indicate differences between the FS of two compounds. Quantum oscillation studies \cite{AWang2016,Wang2018} in YbMnBi$_2$ detected two frequencies with Dirac-like dispersion and large carrier density ($\sim 10^{21}$ cm$^{-3}$) comparable to other Weyl semimetals like Cd$_3$As$_2$ and NbP. In contrast, observation of a single quantum oscillation frequency \cite{Kealhofer2018,Wang2018} and two orders of magnitude diminished carrier density \cite{Kealhofer2018,Wang2018,Pan2021} in YbMnSb$_2$ is difficult to reconcile within existing theoretical results.

In this Letter, we have used neutron and x-ray diffraction to characterize the proper magnetic and crystal structures of YbMnSb$_2$ and investigate the correlation between the crystal/magnetic structures and FS topology via magnetic susceptibility, magnetotransport measurements and first principles calculation. We have identified that YbMnSb$_2$ crystallizes in an orthorhombic $Pnma$ structure with notable anti-phase displacement between the neighboring layers of the magnetic Mn ions. Band structure calculation, utilizing the newly obtained structural parameters, yields a reduced Brillouin zone (BZ) and more anisotropic FS topology of YbMnSb$_2$ than the Bi-based sister compound. 

Earlier studies \cite{Hu2023,Tobin2023,Soh2021,Qiu2019,Kealhofer2018,Wang2018,Pan2021, Baranets2021} reported that YbMnSb$_2$ has a layered tetragonal $P4/nmm$ structure [Fig.~\ref{SCD}(a)]. However, the neutron scattering studies of YbMnSb$_2$ single crystal taken at $T$ = 300 K clearly shows the presence of notable reflections which should be absent in the tetragonal $P4/nmm$ space group \cite{Soh2021}. The origin of those forbidden reflections was not identified, partly due to the limited observable peaks. To reveal the proper crystal structure, we use the white beam neutron diffractometer Corelli covering a large volume in reciprocal space \cite{ye18}. Figs.\ref{SCD}(c) and (d) illustrate the contour map of the neutron scattering pattern obtained at $T$ = 390 K and $T$ = 100 K. Note that Bragg reflections both in the ($H$,$K$,0) and ($H$,0,$L$) planes are observed due to the twinning of the crystal. Reflections at ($2n+1$,0,$m$) position, which corresponds to (integer, 0, half-integer $L$) in the $P4/nmm$ cell, show considerable intensities and clearly indicate the doubling of $c$-axis parameter of the $P4/nmm$ cell. In addition, a close examination of the reflection condition shows difference between equivalent reflections in the tetragonal space group, implying a lowering of the crystal structure symmetry, which is further confirmed by the single crystal x-ray diffraction. Both neutron and x-ray diffraction data can only be modeled and refined using an orthorhombic $Pnma$ structure (space group No.~62) instead of a tetragonal $P4/nmm$ structure. For analysis of structural refinements see the supplemental material (SM) Fig.~S1 and Fig.~S2 \cite{SM}. In the revised structure, Mn, Yb, and Sb ions are located at the $4c$ sites, all showing notable displacement along the $c$-axis [Fig.~1(b)]. The magnitude of the displacement increases as the system is cooled and leads to enhancement of the characteristic $(2n+1,0,$m$)$ reflections.
\begin{figure}[b]
	\includegraphics[width=0.46\textwidth]{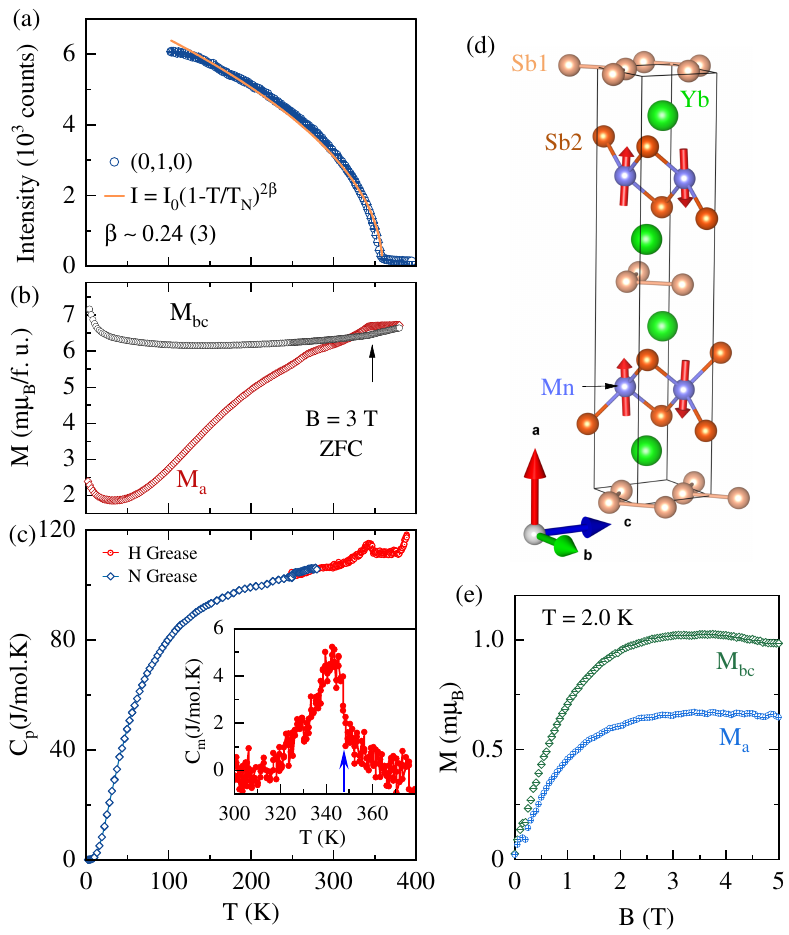}
	\caption{Temperature dependence of (a) the (0,1,0) reflection peak as a magnetic order parameter, (b) zero-field cooled magnetization measured at 3 T, and (c) heat capacity, $C_p$, showing the AFM ordering temperature. Arrows indicate AFM transition temperature. (c) shows background subtracted $C_p$ near the AFM transition. The yellow solid line in (a) represents a fit $I = I_0(1-T/T_{\rm{N}})^{2\beta}$ to the data. $\beta\sim0.24$ indicates a quasi 2D nature of magnetism. (d) The refined spin structure of the YbMnSb$_2$. The spin carries dominant $a$-axis component forming a $C$-type magnetic structure with finite $c$-axis component forming FM sheets coupled antiferromagnetically between the layers. (e) The extracted canted FM moment along the $bc$ plane and the $a$-axis after subtracting the magnetic contribution from FM impurity and AFM order.} 
	\label{canting}
\end{figure}

After establishing the crystal structure, we now discuss the spin arrangement in YbMnSb$_2$. Fig.~\ref{canting}(a) shows the temperature dependence of peak intensity of the $(0, 1, 0)$ magnetic reflection; it decreases sharply upon warming and becomes featureless background above the $T_{\rm{N}}$ $\sim$350 K, consistent with the transition determined from the magnetization and heat capacity [Figs.~\ref{canting}(b)-\ref{canting}(c)]. This implies a spin structure with a unit cell identical to the nuclear one and moments predominately perpendicular to the basal plane. For magnetic ions located at $4c$ site with propagation wavevector $q_m=(0,0,0)$, there are eight compatible magnetic space groups. Half of them can be excluded since the spin moments in those configurations are  constrained within the basal plane, which contradicts the bulk magnetization data. For the remaining magnetic space groups $Pn'm'a'$, $Pnm'a'$, $Pn'm'a$, and $Pnm'a$, the refinement reveals that the magnetic space group $Pn'm'a'$ provides the most satisfactory description of the diffraction data [Fig.~\ref{canting}(d)]. The Mn spin direction lies along the longest crystal axis, the $a$-axis, with a size of $\rm \sim 3.17(3)~\mu_B$ in a collinear $C$-type AFM arrangement. Although the size of the estimated magnetic moment is similar to that previous report \cite{Soh2021}, it is smaller than the expected value for a full ordered Mn$^{2+}$ ($\rm 5~\mu_B$).

Spin canting, which is allowed from the magnetic space group, could be present since finite intensities were observed at $(2n+1, 0,0)$. However, our polarized neutron experiments performed at room temperature revealed that a majority of the intensities originate from the nuclear component [Fig.S3 in SM\cite{SM}]. To estimate the canted moments accurately, magnetization measurements were performed. In Fig.~\ref{canting}(e), we plot the contribution of the canted FM moment to the magnetization after subtracting the contribution of the FM impurity and the AFM ordered state from the magnetization as described in the SM \cite{SM}. The maximum moment of $\sim$0.001 $\mu_B$ is comparable with previous reports in YbMnSb$_2$ \cite{Soh2021} and YbMnBi$_2$ \cite{Pan2022} with a canting angle $\theta\sim$ 0.018$^{\circ}$. This indicates a negligible canting of Mn moment away from the $a$-axis.
\begin{figure}[th]
	\includegraphics[width=0.46\textwidth]{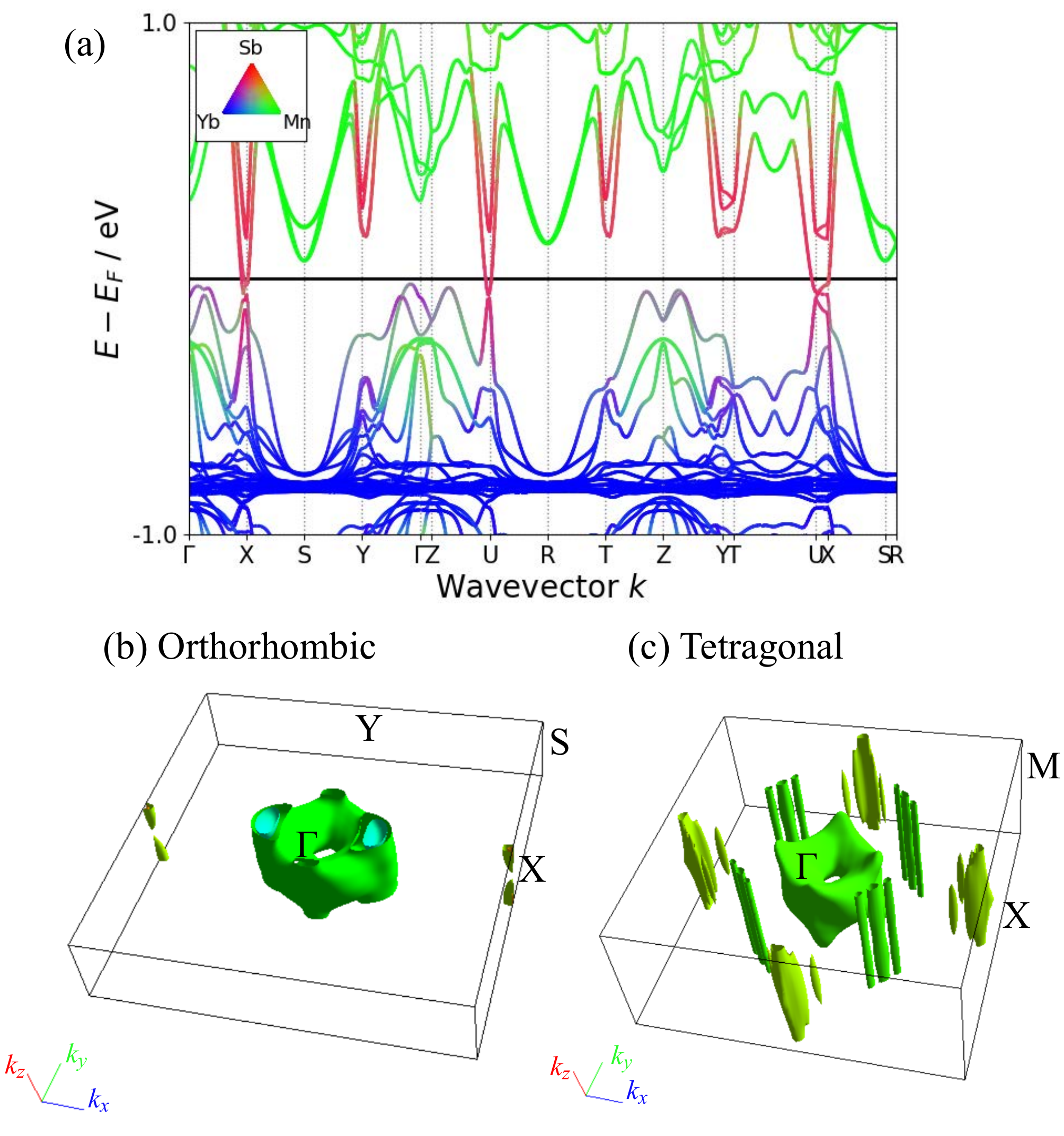}
	\caption{(a) Momentum-dependent electronic structure of YbMnSb$_2$ in the orthorhombic $Pnma$ space group with collinear AFM arrangement of Mn spins. Color represents the orbital contribution of each atom type as shown in the inset. Brillouin zone of YbMnSb$_2$ for (b) orthorhombic and (c) tetragonal structure showing the FS. For tetrgonal structure $E_{\rm{F}}$ is shifted by -50 meV. Due to doubling of unit cell volume, BZ volume in orthorhombic structure becomes half of tetragonal phase.} \label{FS}
\end{figure}

The orthorhombic $Pnma$ space group enforces a zig-zag arrangement of the Sb atoms [as in Fig.~\ref{SCD}(b)] along the $b$-axis leading to a distorted Sb$_1$ layer similar to (Ca/Sr/Ba/Eu)MnSb$_2$ \cite{He2017, Yi2017a, Liu2021, Sakai2020,Gong2020}. Despite the in-plane orthorhombicity $(b-c)/c\sim$ 0.31\% in YbMnSb$_2$ being several times smaller than that in $A$MnSb$_2$ materials, it is sufficient to drive an anisotropic FS compared to a tetragonal structure. In Fig.~\ref{FS}(a), we show the band structure for YbMnSb$_2$ with the collinear AFM arrangements of Mn spin and $Pnma$ space group calculated using density functional theory (see SM for calculation details \cite{SM}) without considering spin-orbit coupling (SOC). The low energy band structure consists of heavier regular bands near the $\Gamma$-point and a linearly dispersing Dirac-like band at the $X$ point. The former arises from Mn $d$-orbitals and Sb $p$-orbitals, whereas the latter mainly originates from the Sb $p$-orbitals. When SOC is taken into account, it has little effect on the bands near the $\Gamma$ point but dramatically increases the gap size at the $X$ point [Fig.~S5 in SM \cite{SM}]. Figs.~\ref{FS}(b) and \ref{FS}(c) compare the BZ of YbMnSb$_2$ in orthorhombic and tetragonal structures, respectively.
The FS in the undistorted tetragonal phase composes of two Dirac-like pockets; one electron-like near the $X$ points and another hole-like along $\Gamma$-$M$ line, and a big 3D hole pocket at the $\Gamma$ point, in good agreement with several previous reports \cite{Pan2022,Pan2021,Le2021,Qiu2019}. However, due to in-plane orthorhombicity, the FS no longer exhibits the $C_4$ rotational symmetry . Moreover, the hole pocket along the $\Gamma$-$M$ direction becomes gapped and the 3D hole pocket at $\Gamma$ stretches along the $\Gamma$-$X$ direction.  

\begin{figure*}[th]
	\includegraphics[width=1.0\textwidth]{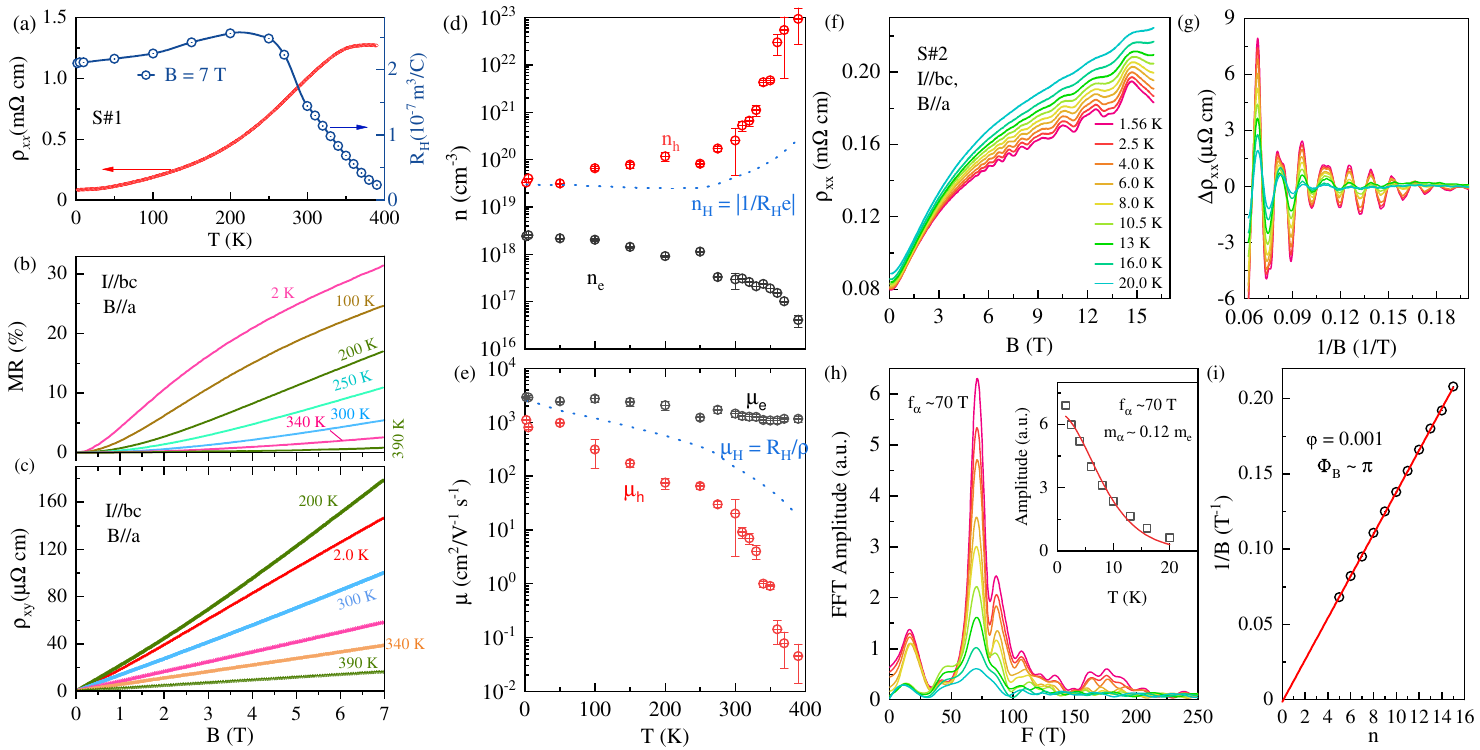}
	\caption{(a) Temperature dependence of $bc$ plane resistivity, $\rho_{xx}$, and Hall coefficient $R_{\rm{H}}$ at 7 T. Magnetic field dependence of (b) magnetoresistance defined as $[(\rho_{xx}(B)-\rho_{xx}(0))/\rho_{xx}(0)]\times100$, and (c) the hall resistivity $\rho_{xy}$ in the temperature range 2 to 390 K. Temperature variation of hole and electron (d) carrier concentration and (e) mobility estimated from the fitting of a two-band model to the MR and hall data. The blue dotted lines represent the carrier concentration, $n_{\rm{H}} = |1/R_He|$, and mobility, $\mu$ = $R_{\rm{H}}$/$\rho_{xx}$, estimated using one band model. (f) Field dependence of $\rho_{xx}$ of crystal $S\#2$ showing SdH quantum oscillations. (g) The corresponding background-subtracted $\Delta$$\rho_{xx}$ vs. $1/B$. (g) FFT spectra of the $\Delta$$\rho_{xx}$ at different temperatures. Inset: LK fit (Equn.\ref{rt}) to the FFT amplitudes of the $f_{\alpha}$. (h) Landau level fan diagram for $f_{\alpha}$.}\label{rho}
\end{figure*}

Fig.~\ref{rho}(a) shows the in-plane resistivity $\rho_{xx}$ and the Hall coefficient $R_{H}$ at 7~T in the temperature range 2~K to 390~K. With decreasing temperature, $\rho_{xx}$ remains nearly flat down to $T_{\rm{N}}$ and decreases sharply below $T_{\rm{N}}$. $R_{\rm{H}}$ increases by an order of magnitude from 2.3$\times$10$^{-8}$ m$^3$/C at 390 K to 2.43$\times$10$^{-7}$ m$^3$/C at 250~K and decreases slightly on further cooling. The estimated carrier concentration $n_{\rm{H}} = |1/R_{\rm{H}}e|$ $\sim$ 2.19 $\times$10$^{-19}$ cm$^{-3}$ at 2 K is comparable with previous reports \cite{Kealhofer2018,Wang2018}, but two orders of magnitude smaller than YbMnBi$_2$ \cite{AWang2016}. To obtain more insight, we measured the magnetic field dependence of $\rho_{xx}$ [Fig.~\ref{rho}(b)] and the Hall resistivity $\rho_{xy}$ [Fig.~\ref{rho}(c)] at representative temperatures across the transition. At low temperatures, MR follows quadratic behavior in the low-field region and saturates at higher fields. For temperatures above 200 K, MR increases quadratically in the whole field region implying the multiple bands at the FS contribute to the charge transport. Consistent with previous results \cite{Wang2018,Kealhofer2018}, $\rho_{xy}$ remains positive revealing that holes are dominant charge carriers. $\rho_{xy}$ follows a linear increase up to 7 T for $\rm 300~K<$ T $< 390$~K but exhibits a concave upward increase below 250 K. Such nonlinear $\rho_{xy}(B)$ indicates that a relatively small number of highly mobile electron-like carriers contribute to the transport property as temperature decreases. 

The magnetic field dependence of $\rho_{xy}$ in a multi-band system is determined by the interplay of concentration and mobility of individual carriers. Hence, we analyzed the corresponding $\rho_{xx}(B)$ and $\rho_{xy}(B)$ employing the semiclassical two-band model as described in SM \cite{SM}. Figs.~\ref{rho}(d)-\ref{rho}(e) show the thermal evolution of the electron (hole) concentration $n_e(n_h)$ and electron (hole) mobility $\mu_e(\mu_h)$ extracted from the two-band model. The concentration of the hole carriers, $n_h$, is an order of magnitude larger than the electron-type carriers, $n_e$, while the mobility of the electron-like carriers, $\mu_e$, is twice that of $\mu_h$. Surprisingly, $n_e$ and $\mu_e$ do not show strong temperature dependence. In contrast, $n_h$ and $\mu_h$ display dramatic temperature dependence as they might be coming from the parabolic bands near the $\Gamma$ points [in Fig.~\ref{FS}(b)]. Both, $n_h$ falls and $\mu_h$ rises by two orders of magnitude across the magnetic transition suggesting that the hole pocket is partially gapped due to band folding as YbMnSb$_2$ transitions from PM to AFM ordered state. 

To further deduce several important physical parameters related to the FS, we measured $\rho_{xx}$ of another piece of crystal $S\#2$ up to 16 T as shown in Fig.~\ref{rho}(f). As the magnetic field exceeds 6 T, prominent Shubnikov de Hass (SdH) quantum oscillations are detected. Fig.~\ref{rho}(g) presents the background subtracted $\Delta$$\rho_{xx}$ \textit{vs.} $1/B$ showing the quantum oscillation at different temperatures up to 20 K. The corresponding fast Fourier transformation (FFT) reveals a primary frequency at $f_{\alpha}\simeq$ 70 T [Fig.~\ref{rho}(h)], in good agreement with previous dHvA and SdH studies of YbMnSb$_2$ \cite{Wang2018}. The frequency of quantum oscillation is related to the cross-section area $S_{F}$ of FS perpendicular to the applied $B$ direction via the Onsager relation, $S_F$ = $(2\pi^{2}/\phi_{0})F$, where $\phi$$_{0}$ is the single magnetic flux quantum. Using this relation, $S_F$ for $f_{\alpha}$ is estimated as 0.007 $\rm \AA^2$, representing a tiny FS cross-sectional area of only 0.3\% of the BZ area $(2\pi/b)\times(2\pi/c)$ = 2.12 \AA$^2$. 

We also analyzed the SdH oscillations quantitatively using the Lifshitz-Kosevich (LK) formula \cite{land, shoe}, which predicts the oscillatory component of $\rho_{xx}$, $\Delta\rho$, as
\begin{equation}
\frac{\Delta\rho}{\rho(0)}\simeq\frac{5}{2}\left(\frac{\mu_0H}{2F}\right)^{\frac{1}{2}}R_{T}(T)R_{D}(T_D)
	 \cos\left[2\pi\left(\frac{F}{B}-\varphi\right)\right],
	\label{lk}
\end{equation}
where $\rho(0)$ is the resistivity $\rho_{xx}$ at $B$ = 0. The cosine term contains a phase factor $\varphi$ = $\frac{1}{2}-\frac{\phi_B}{2\pi}-\delta$, in which $\phi_B$ is the Berry's phase and $\delta$ is related to FS curvature. $\delta$ = 0 for a smooth 2D cylinder, whereas $\delta$ = $\pm$1/8 for a 3D FS. In Eq.(\ref{lk}), the Landau level broadening and electron scattering result in two major damping factors, namely, the temperature damping factor $R_{T}(T)$ and the Dingle factor $R_{D}(T_D)$, respectively:
\begin{equation}
	R_{T}(T)= \frac{2\pi^2k_BTm^{\ast}}{\hbar eB}\sinh\left(\frac{2\pi^2k_BTm^{\ast}}{\hbar eB}\right)
	\label{rt}
\end{equation}
and
\begin{equation}
	R_{D}(T_D) = \exp\left(-\frac{2\pi^2k_BT_Dm^{\ast}}{\hbar eB}\right),
\end{equation}
which are determined by the cyclotron effective mass $m^\ast$ and the Dingle temperature $T_D$. Fitting of Eq.(\ref{rt}) to the thermal damping of FFT amplitude of $f_{\alpha}$ [inset of Fig.~\ref{rho}(h)] results the $m^\ast$$\sim$ 0.12$m_e$, similar to the previous reports \cite{Wang2018, Kealhofer2018}. 

To identify the topological nature of $f_{\alpha}$, the Landau level (LL) fan diagram is employed by plotting the $\Delta\rho_{xx}$ maxima in SdH oscillations against their associated LL index, $n$, in Fig.~\ref{rho}(i). The $x$-intercept of a linear fit to these data provides the accrued $\phi_B$ when the carrier completes one cyclotron orbit, via the relation, $\varphi$ = $1/2-\phi_B/2\pi-\delta$. We assume $\delta$ = 0, as previous SdH oscillation studies have established 2D nature of $f_{\alpha}$ \cite{Wang2018,Kealhofer2018}. $\varphi$ = 0 for a topologically nontrivial Berry’s phase of $\pi$, while a trivial Berry’s phase of 0 results in a $\varphi$ = 1/2. A $\varphi = 0.001(4)$ in the present study indicates that the Fermi pocket giving rise to the SdH oscillations is consistent with having a topological origin and Dirac-like dispersion. 

In summary, we have used neutron and x-ray single crystal diffraction, magnetic susceptibility, and magnetotransport measurements together with complementary band structure calculation to investigate the crystal and magnetic structure as well as the FS topology of YbMnSb$_2$. Both the x-ray and neutron diffraction unambiguously reveal that YbMnSb$_2$ crystallizes in an orthorhombic $Pnma$ structure. Band structure calculation revealed a reduced BZ and more anisotropic FS of YbMnSb$_2$ compared to YbMnBi$_2$ because of in-plane orthorhombicity. The FS of YbMnSb$_2$ consisting of an anisotropic heavier regular band and a linearly dispersing Dirac-like band no longer exhibits the $C_4$ rotational symmetry as in undistorted YbMnBi$_2$. Analysis of SdH quantum oscillation reveals a non-trivial nature of tiny Fermi pocket consistent with Dirac-like energy-momentum dispersion.

\textbf{Acknowledgment:}  We are thankful to S. Nagasaki, D. Hamane, T. Miyake and T. Masuda for technical help during experiments. Also, we gratefully acknowledge fruitful discussion with M. tokunaga, K. Matsuyabashi and P. Shahi. This work was financially supported by the JSPS KAKENHI Grant number JP19H00648. A portion of this research used resources at SNS, a DOE Office of Science User Facility operated by the Oak Ridge National Laboratory. The polarized neutron scattering experiment at JRR-3 was carried out along the proposal No. 22401. The Fermi surfaces and band structure figures were respectively plotted using FermiSurfer \cite{fermisurfer} and pymatgen \cite{Shuye2013}.

\newpage

\title{Supplemental material for \textquotedblleft Fermi surface reconstruction due to the orthorhombic distortion in Dirac semimetal YbMnSb$_2$\textquotedblright}
\author{Dilip Bhoi}
\affiliation{The Institute for Solid State Physics, University of Tokyo, Kashiwa, Chiba 277-8581, Japan}
\author{Feng Ye}
\affiliation{Neutron Scattering Division, Oak Ridge National Laboratory, Oak Ridge, Tennessee 37831, USA}
\author{Hanming Ma}
\author{Xiaoling Shen}
\affiliation{The Institute for Solid State Physics, University of Tokyo, Kashiwa, Chiba 277-8581, Japan}
\author{Arvind Maurya}
\affiliation{Department of Physics, School of Physical Sciences, Mizoram University, Aizawl 796 004, India}
\author{Shusuke Kasamatsu}
\affiliation{Academic Assembly (Faculty of Science), Yamagata University, Yamagata 990-8560, Japan}
\author{Takahiro Misawa}
\author{Kazuyoshi Yoshimi}
\author{Taro Nakajima}
\affiliation{The Institute for Solid State Physics, University of Tokyo, Kashiwa, Chiba 277-8581, Japan}
\author{Masaaki Matsuda}
\affiliation{Neutron Scattering Division, Oak Ridge National Laboratory, Oak Ridge, Tennessee 37831, USA}
\author{Yoshiya Uwatoko}
\affiliation{The Institute for Solid State Physics, University of Tokyo, Kashiwa, Chiba 277-8581, Japan}
\date{\today}
\newpage
\newpage
\section{Supplemental Material}
\setcounter{figure}{0}
\subsection*{Single crystal growth, experimental methods and techniques}
We have prepared large single crystals of YbMnSb$_2$ using antimony as a flux. The high-purity elements of Yb, Mn, and Sb with a 1:1:4 stoichiometric ratio were placed in an alumina crucible as starting materials. To grow the single crystals we followed the recipe as described in Ref. \cite{Wang2018}. The as-grown single crystals were extensively characterized by single crystal x-ray and neutron diffraction at room temperature and energy dispersive x-ray analysis (EDX). To check the chemical homogeneity of the crystal, EDX spectra obtained at several randomly selected spots from the samples were analyzed. The analyses show that the crystals are chemically homogeneous within the limit of EDX with an average stoichiometric ratio, Yb: Mn: Sb = 0.98:1.01:2.01. 

A rectangular piece of YbMnSb$_2$ single crystal was cut for magnetic susceptibility measurement in a Superconducting Quantum Interference Device (MPMS XL, Quantum Design). The electrical resistivity and heat capacity measurements in the temperature range between 2 to 390 K were performed using a Quantum Design Physical property measurement system. For Shubnikov-de Hass (SdH) quantum oscillation measurements up to 16 T an oxford superconducting magnet is used. 

Neutron diffraction measurements were performed on the SNS CORELLI Beamline-09 installed at Oak Ridge National Laboratory, USA using a piece of single crystal with a weight of $\sim$ 100 mg. The single crystal was cooled down using a closed-cycle He4 refrigerator. Data reduction and analysis were conducted using MANTID software \cite{Clark2916}. Magnetic structures were analyzed with SARAH software \cite{Willis2000}. All the refinements of the neutron scattering data were performed with FULLPROF software \cite{Carvajal1993}.
\begin{figure}[th]
\renewcommand{\figurename}{Fig.S}
\includegraphics[width=0.34\textwidth]{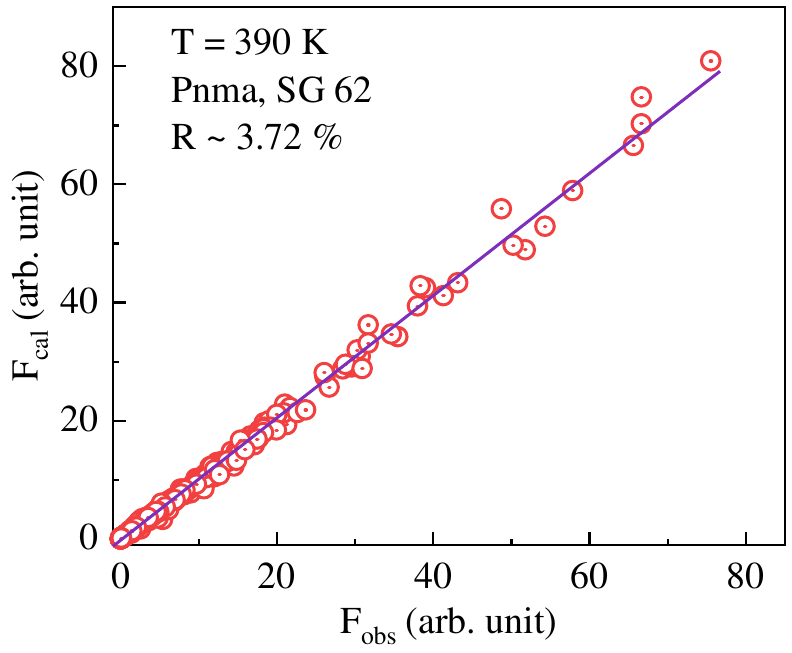}
\caption{The results of refinements of the neutron scattering intensity at $T$ = 390 K with the orthorhombic structure and $Pnma$ space group no. 62.}\label{NRef}
\end{figure}
\begin{table}[ht]
\caption{\label{CSP}{Refined structural parameters for YbMnSb$_2$ from neutron diffraction at 390 K in the orthorhombic $Pnma$ space group, with $a$ = 21.5963(8) \AA, $b$ = 4.3235(17) \AA, $c$ =  4.3099(13) \AA, $\alpha$ = $\beta$ = $\gamma$ = 90$^{\circ}$. Unit cell volume $V$ = 402.43(2) \AA$^3$, Number of formula unit per unit cell $Z$ = 4.}} 
\begin{ruledtabular}
\begin{tabular}{ c c c c c c}
Atoms & Wyckoff position & x  &  y &  z\\[0.5ex]
\hline
Yb  & 4c & 0.38612(4) & 1/4 & 0.2629(2)\\ 
Sb1 & 4c& 0.00099(9) & 3/4 & 0.2677(3) \\ 
Sb2 & 4c& 0.32985(7) & 3/4 & 0.7627(2) \\ 
Mn  & 4c& 0.2498(3) & 3/4 & 0.2646(6)\\
\end{tabular}
\end{ruledtabular}
\end{table}
\subsection{Single crystal neutron and x-ray diffraction}
\begin{figure*}[th]
\includegraphics[width=0.9\textwidth]{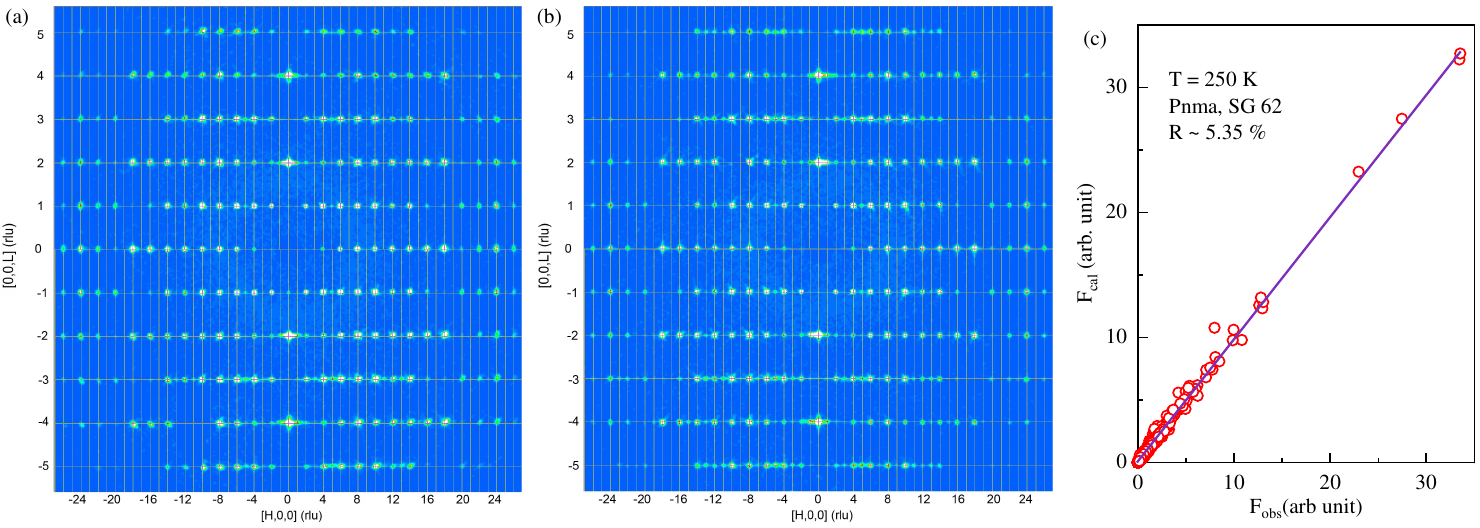}
\renewcommand{\figurename}{Fig.S}
\caption{Image of single crystal x-ray diffraction patterns of YbMnSb$_2$ at $T$= 250 K (a) in the ($H$,$K$,0) plane satisfying the reflection condition $H$ = 2$n$ for the $Pnma$ space group, and (b) in the ($H$,0,$L$) plane confirming the doubling of unit cell along the $a$-axis. (c) The refinements of the x-ray diffraction pattern at $T$ = 250 K with the orthorhombic structure and $Pnma$ space group no. 62.}\label{xray}
\end{figure*}
The neutron scattering pattern at $T$ = 390 K is successfully refined assuming the orthorhombic structure with $Pnma$ space group no.62 as evidenced by the good agreement between calculated structure factor square ($F^2_{cal}$) and observed structure factor square ($F^2_{obs}$) [in Fig.S\ref{NRef}]. To further confirm, we checked the single crystal x-ray diffraction of YbMnSb$_2$ (with a typical crystal size $\sim$ 60 $\mu$m) at $T$ = 250 K. Figs.S\ref{xray}(a) and (b) show the single crystal x-ray diffraction patterns in the ($H$,$K$,0) and ($H$,0,$L$) reciprocal planes, respectively. Reflections in the former plane satisfies the necessary reflection condition $H$ = 2$n$ for the $Pnma$ space group No.62. Note that the sample used in this x-ray experiment was nearly single domained. Therefore, the ($H$,0,$L$) reflections from the other twin were weak [Fig.S\ref{xray}]. The refinement of the x-ray diffraction indeed ascertain that YbMnSb$_2$ crystallizes in orthorhombic $Pnma$ structure rather than the tetragonal centrosymmetric $P4/nmm$ suggested in previous reports \cite{Soh2021, Kealhofer2018, Wang2018}. As can be seen in Fig.S\ref{xray}(c), the $F^2_{cal}$ matches very well with the $F^2_{obs}$, indicating excellent structure refinement. The details of the refined crystal structural parameters are given in Table.\ref{CSP}. 
\subsection{Polarized neutron scattering}
\begin{figure}[h]
\includegraphics[width=0.5\textwidth]{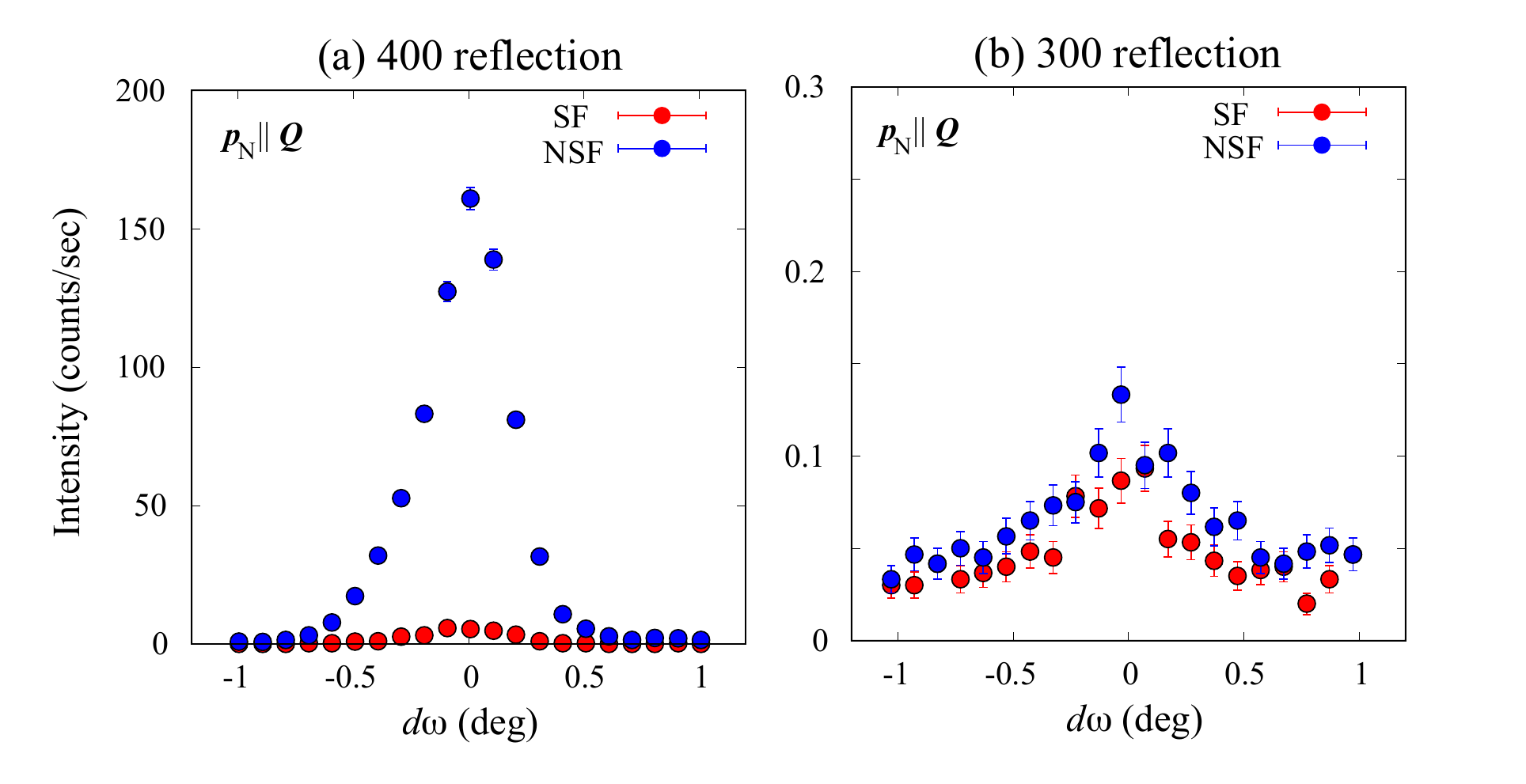}
\renewcommand{\figurename}{Fig.S}
\caption{Polarized neutron scattering profiles of (a) (4,0,0) and (b) (3,0,0) reflections measured at room temperature. The beam polarization was 0.928.}\label{sx}
\end{figure}
We performed polarized neutron scattering measurements at the polarized neutron triple-axis spectrometer (PONTA) installed in JRR-3. The spectrometer was operated in the $P_{xx}$ longitudinal polarization analysis mode, in which the direction of the neutron spin polarization, $p_N$, was set to be parallel or antiparallel to the scattering vector, $Q$ (=$k_i-k_f$) by guide fields and a spin flipper. We measured intensities of spin-flip (SF) and non-spin-flip (NSF) scattering processes, in which the spins of the incident neutrons were flipped and remained unchanged, respectively. In the present setup, magnetic and nuclear scattering signals should be separated from each other, and observed in the SF and NSF channels, respectively. Figs.S\ref{sx}(a) and (b) show the rocking curve profiles of the (4,0,0) and (3,0,0) reflections, respectively, at room temperature. As shown in Fig.S\ref{sx}(a), the NSF scattering dominate over the SF scattering at the (4,0,0) reflection, which confirms that the (4,0,0) fundamental reflection does not contain magnetic components. Very weak intensities in the SF channel are due to the imperfect polarization of the incident neutron beam. We found that the NSF scattering was also stronger in the (3,0,0) superlattice reflection as shown in Fig.S\ref{sx}(b), suggesting that the origin of this superlattice reflection is mainly attributed to the nuclear component. Note that the intensities at (3,0,0) contain a very small half-lambda contamination, which equally contributes to both the SF and NSF intensities. 

\subsection{Spin canting moment from magnetization}
\begin{figure}[th]
\includegraphics[width=0.35\textwidth]{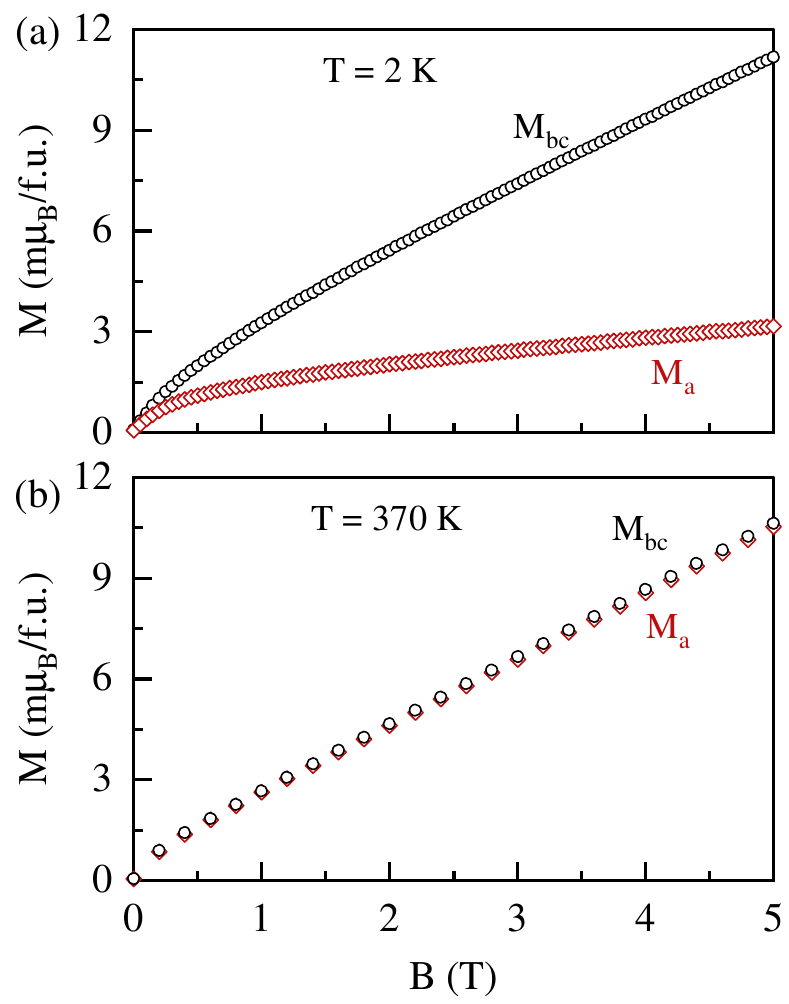}
\renewcommand{\figurename}{Fig.S}
\caption{Magnetic field dependence of $M_{bc}$ and $M_{a}$ at (a) $T$ = 2 K in the AFM ordered state, and (c) at $T$ = 370 K in the paramagnetic state.}\label{mag}
\end{figure}

\begin{figure}[th]
\includegraphics[width=0.3\textwidth]{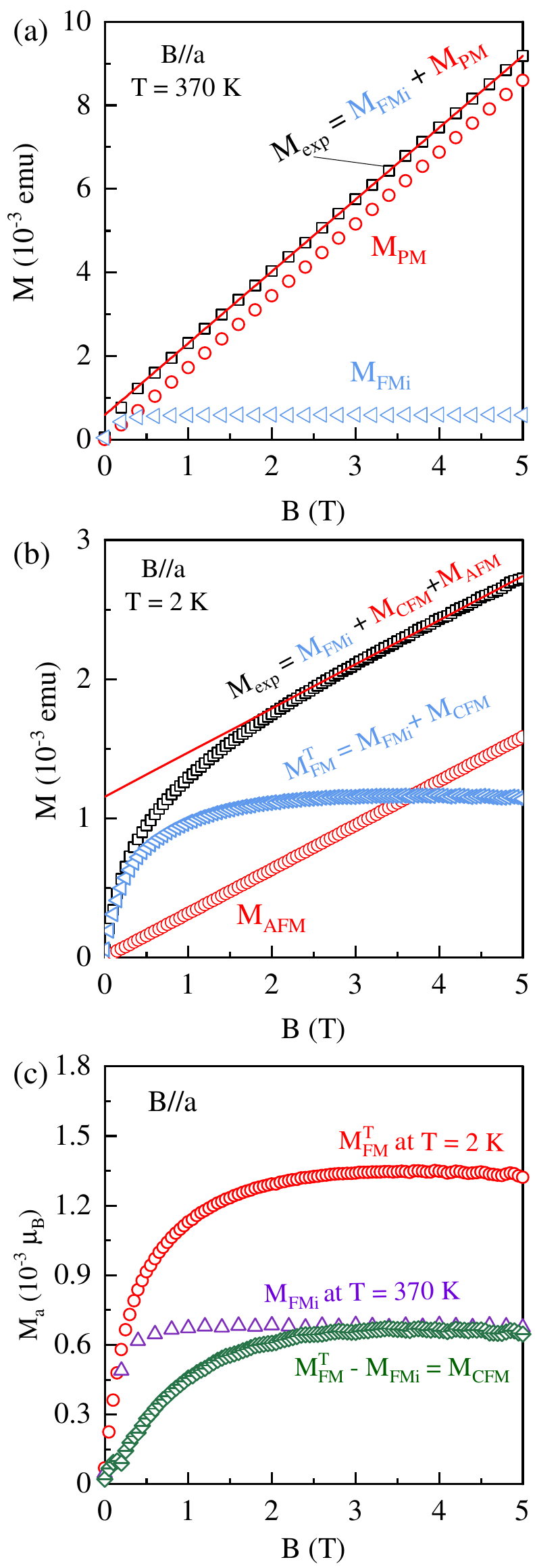}
\renewcommand{\figurename}{Fig.S}
\caption{The field dependence of the magnetization at (a) $T$ = 370 K and (b) $T$ = 2 K for magnetic field applied parallel to $a$-axis. The red lines represent the linear fit to the data at high field region. The black square ($\Box$), red circle (\textcolor{red}{$\bigcirc$}) and blue triangle (\textcolor{blue}{$\lhd$}) represent the experimental data, $M_{\rm{exp}}$, PM/AFM contribution, $M_{\rm{PM/AFM}}$, and the contribution from FM impurity, $M_{\rm{FMi}}$ or $M_{\rm{FMi}}^{\rm{T}}$, respectively. (c) The extracted FM contribution arising from the canted moment after subtracting the magnetization from the FM impurity and AFM order at $T$ = 2 K.}\label{cmag}
\end{figure}

In Fig.2(b) of the main text, we show the temperature dependence of the magnetization $M_{bc}$ and $M_{a}$  of the YbMnSb$_2$ measured at $H$ = 3 T, respectively. The temperature dependence of magnetizations and the estimated transition temperature from the temperature dependence of the magnetization $M_{bc}$ and $M_{a}$ are consistent with previous reports \cite{Soh2021,Wang2018}. With decreasing temperature, $M_{bc}$ and $M_{a}$ bifurcates below $T_{\rm{N}}$$\sim$345 K; $M_{a}$ decrease rapidly, whereas $M_{bc}$ reveals weak temperature dependence. Below 30 K, both $M_{bc}$ and $M_{a}$ reveal a weak upturn as temperature decreases. Fig.S\ref{mag}(a) and (b) show the field dependence of the $M_{bc}$ and $M_{a}$ at $T$ = 2.0 K and $T$ = 370 K, respectively. In the PM state at $T$ = 370 K, both $M_{bc}$ and $M_{a}$ clearly overlap on top of each other, whereas below $T_{\rm{N}}$, the difference between $M_{bc}$ and $M_{a}$ start to grow with decreasing temperature. The persistence of FM signal below $B<$ 0.5 T even in the PM state implies that the weak upturn in $M_{bc}$ and $M_{a}$ at low temperatures is likely to arise from a trace amount of MnSb$_2$ impurity in the sample. \\
To extract the size of intrinsic magnetic moment arising from the canting of spin, we followed a method as explained below. For simplicity, we only describe the procedure applied for $B\parallel a$-axis. Figs.S\ref{cmag}(a) and (b) show the field dependence of the magnetization of YbMnSb$_2$ at $T$ = 370 K and $T$ = 2 K, respectively. The experimental data $M_{\rm{exp}}$ in PM state ($T$ = 370 K) can be assumed as the sum of the magnetization arising from the FM impurity, $M_{\rm{FMi}}$, and the PM state, $M_{\rm{PM}}$ as shown in Fig.S\ref{cmag}(a). The magnetic contribution from the $M_{\rm{FMi}}$ is estimated by subtracting the $M_{\rm{PM}}$ from $M_{\rm{exp}}$ following a similar method used to extract the contribution of FM impurity in iron based single crystals \cite{Johnston}. To calculate the $M_{\rm{PM}}$, as shown in Fig.S\ref{cmag}(a), a linear fit to the $M_{\rm{exp}}$ data in the field range 1 T to 5 T  is applied and extrapolated to zero. By making the finite intercept to zero, the linear line passing through the origin can be assumed as the magnetization due to $M_{\rm{PM}}$. 

However, in the AFM ordered state, in addition to $M_{\rm{FMi}}$, the $M_{\rm{exp}}$ will also contain contributions from the AFM ordered state ($M_{\rm{AFM}}$) and the canted FM component ($M_{\rm{CFM}}$). Therefore, to estimate the $M_{\rm{CFM}}$ a two step procedure is followed. First, following a similar method used to extract $M_{\rm{PM}}$ at $T>T_{\rm{N}}$, $M_{\rm{AFM}}$ at 2 K is extracted by performing the linear fit in the field range 2.5 T to 5 T. After subtraction of $M_{\rm{AFM}}$ from $M_{\rm{exp}}$, the residue will contain the total FM component of magnetization, $M_{\rm{FM}}^{\rm{T}}$ arising from the $M_{\rm{CFM}}$ and $M_{\rm{FMi}}$ as shown in Fig.S\ref{cmag}(b). As $M_{\rm{FMi}}$ is expected to saturate with decreasing temperature, the estimated $M_{\rm{FMi}}$ at $T$ = 370 K provides a good approximation of the $M_{\rm{FMi}}$ even at lower temperatures. Therefore, in the second step, we extract the intrinsic $M_{\rm{CFM}}$, by subtracting the $M_{\rm{FMi}}$ at $T$ = 370 K from the $M_{\rm{FM}}^{\rm{T}}$ as shown in Fig.S\ref{cmag}(c). A similar strategy is followed to estimate the spin canting moment for magnetic field along the $bc$ plane.
\subsection{Electronic band structure}
\begin{figure}[th]
\includegraphics[width=0.45\textwidth]{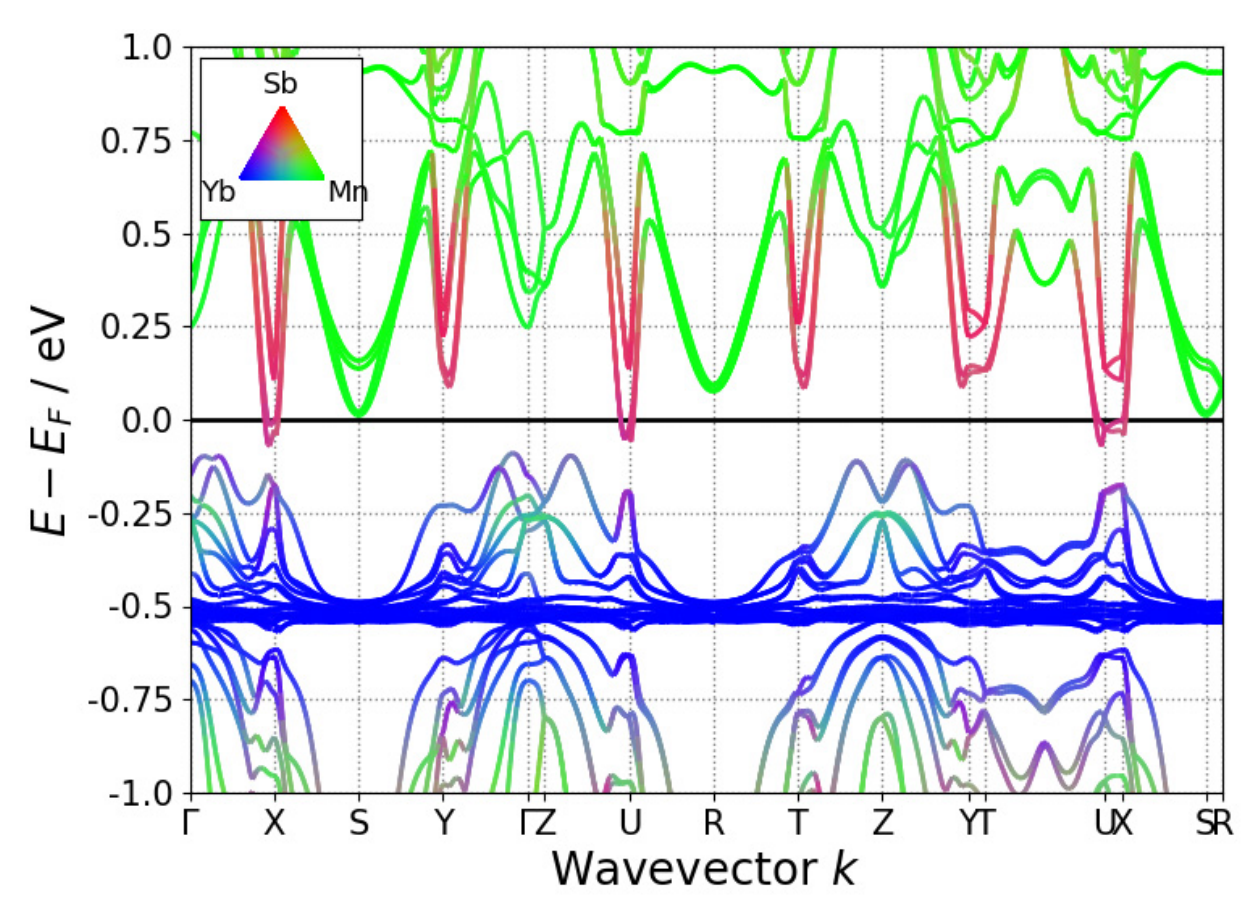}
\renewcommand{\figurename}{Fig.S}
\caption{Electronic band structure of YbMnSb$_2$ with the orthorhombic structure and the collinear C-type AFM arrangement of Mn spins including the spin orbit coupling.}\label{bs}
\end{figure}

The density functional theory calculations for the band structure and Fermi surfaces (Fig.~3 in the main text) were performed using VASP code \cite{Kresse1996} using the GGA-PBE approximation \cite{Perdew1996} with the structure fixed to that determined experimentally in this work. The projector-augmented wave method \cite{Bloechl1994} was employed to describe ion-electron interactions. The energy cutoff was set to 600 eV for the plane wave expansion of the wave functions, and a $2 \times 10 \times 10$ $k$-point mesh was used for the Brillouin zone integration for the self consistent calculations. The rotationally invariant DFT$+U$ approach of Dudarev et al.\cite{Dudarev1998} was used with $U_\text{f}=4$ eV for Yb to reproduce photoemission spectra as per Ref.~\cite{Kealhofer2018}. A non-self consistent calculation with fixed electronic density was performed to obtain eigenvalues along high-symmetry lines for the band structure and on a dense uniform mesh for the Fermi surface calculations. 
Fig.S\ref{bs} in this supplement shows the theory calculated band structure for YbMnSb$_2$ with spin-orbit coupling in the collinear AFM arrangements of Mn spin, while Fig.~3 in the main text was obtained from calculations without spin-orbit coupling.

\subsection{Analysis of magnetoresistance using a two-band model}
\begin{figure}[bh]
\includegraphics[width=0.45\textwidth]{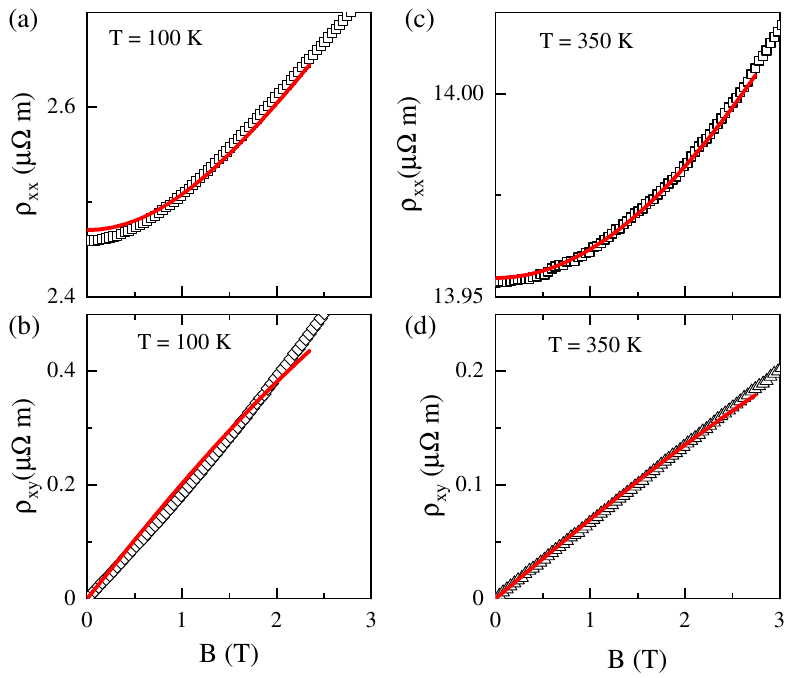}
\renewcommand{\figurename}{Fig.S}
\caption{(a) and (b) shows the simultaneous fitting of Equn.\ref{MR} and Equn.\ref{hall} to the $\rho_{xx}$ and $\rho_{xy}$ data at $T$ = 100 K. The fitting for $\rho_{xx}$ and $\rho_{xy}$ at $T$ =350 K are shown in (c) and (d), respectively}\label{two band}
\end{figure}

To extract the carrier concentrations and mobilities of individual bands, we analyzed the magnetic field dependence of $\rho_{xx}$ and $\rho_{xy}$ at different temperatures employing the semiclassical two-band model. Based on this model $\rho_{xx}(B)$ and $\rho_{xy}(B)$ are described as
\begin{equation}
  \rho_{xx}(B) = \frac{1}{e}\frac{(n_h\mu_h+n_e\mu_e)+n_h\mu_e(n_h\mu_e+n_e\mu_h)B^2}{(n_h\mu_h+n_e\mu_e)^2+\mu_h^2\mu_e^2(n_h-n_e)B^2}\label{MR}
\end{equation}
and 
\begin{equation}
  \rho_{xy}(B) = \frac{1}{e}\frac{(n_h\mu_h^2-n_e\mu_e^2)+\mu_h^2\mu_e^2(n_h-n_e)B^2}{(n_h\mu_h+n_e\mu_e)^2+\mu_h^2\mu_e^2(n_h-n_e)B^2}B, \label{hall}
\end{equation}
where $n_e(n_h)$ and $\mu_e(\mu_h)$ correspond to electron (hole) concentration and electron (hole) mobility, respectively. In Figs.S\ref{two band} (a)-(d), we show the typical examples of simultaneous fitting of Equn.\ref{MR} and Equn.\ref{hall} to $\rho_{xx}(B)$ and $\rho_{xy}(B)$, where they show nonlinear field dependence, respectively.  

\begin{thebibliography}{99}

\bibitem{Wang2011} Kefeng Wang, D. Graf, Hechang Lei, S. W. Tozer, and C. Petrovic, Quantum transport of two-dimensional Dirac fermions in SrMnBi$_2$. Phys. Rev. B \textbf{84}, 220401(R) (2011).

\bibitem{Wang2012} Kefeng Wang, D. Graf, Limin Wang, Hechang Lei, S. W. Tozer, and C. Petrovic, Two-dimensional Dirac fermions and quantum magnetoresistance in CaMnBi$_2$, Phys. Rev. B \textbf{85}, 041101(R) (2012).

\bibitem{Park2011} Joonbum Park, G. Lee, F. Wolff-Fabris, Y. Y. Koh, M. J. Eom, Y. K. Kim, M. A. Farhan, Y. J. Jo, C. Kim, J. H. Shim, and J. S. Kim, Anisotropic Dirac Fermions in a Bi Square Net of SrMnBi$_2$, Phys. Rev. Lett. \textbf{107}, 126402 (2011). 

\bibitem{Jo2014} Y. J. Jo, Joonbum Park, G. Lee, Man Jin Eom, E. S. Choi, Ji Hoon Shim, W. Kang, and Jun Sung Kim, Valley-Polarized Interlayer Conduction of Anisotropic Dirac Fermions in SrMnBi$_2$, Phys. Rev. Lett. \textbf{113}, 156602 (2014).

\bibitem{Liu2017a} J. Y. Liu, J.Hu, D. Graf, T. Zou, M. Zhu, Y. Shi, S. Che, S.M.A. Radmanesh, C.N. Lau, L. Spinu, H.B. Cao, X.Ke, and Z.Q. Mao, Unusual interlayer quantum transport behavior caused by the zeroth Landau level in YbMnBi$_2$. Nat. Commun., \textbf{8(1)}, 646, (2017).

\bibitem{AWang2016} A. Wang, I. Zaliznyak, W. Ren, Lijun Wu, D. Graf, V. O. Garlea, J. B. Warren, E. Bozin, Y. Zhu, and C. Petrovic, Magnetotransport study of Dirac fermions in YbMnBi$_2$ antiferromagnet, Phys. Rev. B, \textbf{94}, 165161 (2016)

\bibitem{Fang2014} Ya Feng, Zhijun Wang, Chaoyu Chen, Youguo Shi, Zhuojin Xie, Hemian Yi, Aiji Liang, Shaolong He, Junfeng He, Yingying Peng, Xu Liu, Yan Liu, Lin Zhao, Guodong Liu, Xiaoli Dong, Jun Zhang, Chuangtian Chen, Zuyan Xu, Xi Dai, Zhong Fang and X. J. Zhou, Strong Anisotropy of Dirac Cones in SrMnBi$_2$ and CaMnBi$_2$ Revealed by Angle-Resolved Photoemission Spectroscopy, Sci. Rep.,\textbf{4}, 5385 (2016).

\bibitem{Liu2017b} J. Y. Liu, J. Hu, Q. Zhang, D. Graf, H. B. Cao, S. M. A. Radmanesh, D. J. Adams, Y. L. Zhu, G. F. Cheng, X. Liu, W. A. Phelan, J.Wei, M. Jaime, F. Balakirev, D. A. Tennant, J. F. DiTusa, I. Chiorescu, L. Spinu, and Z. Q. Mao, A magnetic topological semimetal Sr$_{1-y}$Mn$_{1-z}$Sb$_2$ ($y$, $z <$0.1), Nat. Mater., \textbf{16}, 905 (2017).

\bibitem{He2017} J. B. He, Y. Fu, L. X. Zhao, H. Liang, D. Chen, Y. M. Leng, X. M. Wang, J. Li, S. Zhang, M. Q. Xue, C. H. Li, P. Zhang, Z. A. Ren, and G. F. Chen, Quasi-two-dimensional massless Dirac fermions in CaMnSb$_2$, Phys. Rev. B, \textbf{95}, 045128 (2017).

\bibitem{Zhang2016} Anmin Zhang, Changle Liu, Changjiang Yi, Guihua Zhao, Tian-long Xia, Jianting Ji, Youguo Shi, Rong Yu, Xiaoqun Wang, Changfeng Chen and Qingming Zhang, Interplay of Dirac electrons and magnetism in CaMnBi$_2$ and SrMnBi$_2$, Nat. Commun., \textbf{7}, 13833 (2016).

\bibitem{Huang2017} S. Huang, J. Kim, W. A. Shelton, E. W. Plummer, and R. Jin, Nontrivial Berry phase in magnetic BaMnSb$_2$ semimetal, Proc. Natl. Acad. Sci. USA, \textbf{114} 6256, (2017). 

\bibitem{Liu2016a} Jinyu Liu, Jin Hu, Huibo Cao, Yanglin Zhu, Alyssa Chuang, D. Graf, D. J. Adams, S. M. A. Radmanesh, L. Spinu, I. Chiorescu, and Zhiqiang Mao, Nearly massless Dirac fermions hosted by Sb square net in BaMnSb$_2$, Sci Rep \textbf{6}, 30525 (2016).

\bibitem{Yi2017a} Changjiang Yi, Shuai Yang, Meng Yang, Le Wang, Yoshitaka Matsushita, Shanshan Miao, Yuanyuan Jiao, Jinguang Cheng, Yongqing Li, Kazunari Yamaura, Youguo Shi, and Jianlin Luo, Large negative magnetoresistance of a nearly Dirac material: Layered antimonide EuMnSb$_2$, Phys. Rev. B \textbf{96}, 205103 (2017). 

\bibitem{May2014} A. F. May, M. A. McGuire, and B. C. Sales, Effect of Eu magnetism on the electronic properties of the candidate Dirac material EuMnBi$_2$, Phys. Rev. B \textbf{90}, 075109 (2014).

\bibitem{Borisenko2019} S. Borisenko, D. Evtushinsky, Q. Gibson, A. Yaresko, K. Koepernik, T. Kim, M. Ali, J. van den Brink, M. Hoesch, A. Fedorov, E. Haubold, Y. Kushnirenko, I. Soldatov, R. Sch\"{a}fer and R. J. Cava, Time-reversal symmetry breaking type-II Weyl state in YbMnBi$_2$, Nat. Commun., \textbf{10}, 3424 (2019). 

\bibitem{Masuda2016} H. Masuda, H. Sakai, M. Tokunaga, Y. Yamasaki, A. Miyake, J. Shiogai, S. Nakamura, S. Awaji, A. Tsukazaki, H. Nakao, Y. Murakami, T. Arima, Y. Tokura, S. Ishiwata1, Quantum Hall effect in a bulk antiferromagnet EuMnBi$_2$ with magnetically confined two-dimensional Dirac fermions. Sci. Adv., \textbf{2}, e1501117 (2016).

\bibitem{Liu2021} J. Y. Liu, J. Yu, J. L. Ning, H. M. Yi, L. Miao, L. J. Min, Y. F. Zhao, W. Ning, K. A. Lopez, Y. L. Zhu, \textit{et al.}, Spin-valley locking and bulk quantum Hall effect in a noncentrosymmetric Dirac semimetal BaMnSb$_2$, Nat. Commun., \textbf{12}, 4062 (2021)

\bibitem{Sakai2020} H. Sakai, H. Fujimura, S. Sakuragi, M. Ochi, R. Kurihara, A. Miyake, M. Tokunaga, T. Kojima, D. Hashizume, T. Muro, \textit{et al.}, Phys. Rev. B \textbf{101}, 081104(R) (2020). 

\bibitem{Pan2022} Yu Pan, Congcong Le, Bin He, Sarah J. Watzman, Mengyu Yao, Johannes Gooth, Joseph P. Heremans, Yan Sun  and Claudia Felser, Giant anomalous Nernst signal in the antiferromagnet YbMnBi$_2$, Nat. Mater., \textbf{21}, 203 (2022). 

\bibitem{Le2021} Congcong Le, Claudia Felser, and Yan Sun, Design strong anomalous Hall effect via spin canting in antiferromagnetic nodal line materials, Phys. Rev. B, \textbf{104}, 125145 (2021). 

\bibitem{Li2019arxiv} Xiao Li, Allan H. MacDonald, Hua Chen, Quantum Anomalous Hall Effect through Canted Antiferromagnetism, arXiv:1902.10650v1 

\bibitem{Guo2022arxiv} Peng-Jie Guo, Zheng-Xin Liu, Zhong-Yi Lu, Quantum Anomalous Hall Effect in Antiferromagnetism, arXiv:2205.06702v1

\bibitem{Ni2022} Xiao-Sheng Ni, Cui-Qun Chen, Dao-Xin Yao, and Yusheng Hou, Origin of the type-II Weyl state in topological antiferromagnetic YbMnBi$_2$, Phys. Rev. B, \textbf{105}, 134406 (2022).

\bibitem{Yang2020} R. Yang, M. Corasaniti, C. C. Le, Z. Y. Liao, A. F. Wang, Q. Du, C. Petrovic, X. G. Qiu, J. P. Hu, and L. Degiorgi, Spin-Canting-Induced Band Reconstruction in the Dirac Material Ca$_{1-x}$Na$_x$MnBi$_2$, Phys. Rev. Lett. \textbf{124}, 137201 (2020). 

\bibitem{Saptoka2020} A. Sapkota, L. Classen, M. B. Stone, A. T. Savici, V. O. Garlea, Aifeng Wang, J. M. Tranquada, C. Petrovic, and I. A. Zaliznyak, Signatures of coupling between spin waves and Dirac fermions in YbMnBi$_2$. Phys. Rev. B, \textbf{101}, 041111(R) (2020).

\bibitem{Soh2019} J. -R. Soh, H. Jacobsen, B. Ouladdiaf, A. Ivanov, A. Piovano, T. Tejsner, Z. Feng, H. Wang, H. Su, Y. Guo, Y. Shi, and A. T. Boothroyd, Magnetic structure and excitations of the topological semimetal YbMnBi$_2$. Phys. Rev. B, \textbf{100}, 144431 (2019).

\bibitem{Hu2023} X. Hu, A. Sapkota, Z. Hu, A. T. Savici, A. I. Kolesnikov, J. M. Tranquada, C. Petrovic, and I. A. Zaliznyak, Coupling of magnetism and Dirac fermions in YbMnSb$_2$, Phys. Rev. B. \textbf{107}, L201117 (2023).

\bibitem{Tobin2023} S. M. Tobin, J.-R. Soh, H. Su, A. Piovano, A. Stunault, J. A. Rodr\'{i}guez-Velamaz\'{a}n, Y. Guo, and A. T. Boothroyd, Magnetic excitations in the topological semimetal YbMnSb$_2$, Phys. Rev. B. \textbf{107}, 195146 (2023).

\bibitem{Kealhofer2018}	R. Kealhofer, S. Jang, S. M. Griffin, C. John, K. A. Benavides, S. Doyle, T. Helm, P. J. W. Moll, J. B. Neaton, J. Y. Chan, J. D. Denlinger, and J. G. Analytis, Observation of a two-dimensional Fermi surface and Dirac dispersion in YbMnSb$_2$, Phys. Rev. B. \textbf{97}, 045109 (2018). 

\bibitem{Qiu2019} Ziyang Qiu, Congcong Le, Zhiyu Liao, Bing Xu, Run Yang, Jiangping Hu, Yaomin Dai, and Xianggang Qiu, Observation of a topological nodal-line semimetal in YbMnSb$_2$ through optical spectroscopy, Phys. Rev. B \textbf{100}, 125136 (2019)

\bibitem{Wang2018} Y.-Y. Wang, Yi-Yan Wang, Sheng Xu, Lin-Lin Sun, and Tian-Long Xia, Quantum oscillations and coherent interlayer transport in a new topological Dirac semimetal candidate YbMnSb$_2$, Phys. Rev. Mater., \textbf{2}, 021201(R) (2018). 

\bibitem{Pan2021} Y. Pan, F.-R. Fan, X. Hong, B. He, C. Le, W. Schnelle, Y. He, K. Imasato, H. Borrmann, C. Hess, B. Büchner, Y. Sun, C. Fu, G. Jeffrey Snyder, and C. Felser, Thermoelectric Properties of Novel Semimetals: A Case Study of YbMnSb$_2$, Adv. Mater. \textbf{33}, 2003168 (2021).

\bibitem{Baranets2021} S. Baranets and S. Bobev, Transport properties and thermal behavior of YbMnSb$_2$ semimetal above room temperature. Journal of Solid State Chemistry, \textbf{303}, 122467 (2021). 

\bibitem{Soh2021} J.-R. Soh, S. M. Tobin, Hao Su, I. Zivkovic, B. Ouladdiaf, A. Stunault, J. Alberto Rodríguez-Velamazán, K. Beauvois, Y. Guo, and A. T. Boothroyd, Magnetic structure of the topological semimetal YbMnSb$_2$, Phys. Rev. B, \textbf{104}, L161103 (2021).

\bibitem{ye18} F. Ye, Y. Liu, R. Whitfield, R. Osborn, and S. Rosenkranz, {J. Appl. Crystallogr. {\bf 51}, 315} (2018).

\bibitem{SM} See Supplemental Material for details pertaining to sample synthesis, experimental methods and techniques, Rietveld refinements, polarized neutron scattering, DFT result with SOC, estimation of the spin canting moment, and analysis of magnetotransport data using semiclassical two-band model.

\bibitem{Gong2020} Dongliang Gong, Silu Huang, Feng Ye, Xin Gui, Jiandi Zhang, Weiwei Xie, and Rongying Jin, Canted Eu magnetic structure in EuMnSb$_2$, Phys. Rev. B, \textbf{101}, 224422 (2020). 

\bibitem{Clark2916} T. M. Michels-Clark, A. T. Savici, V. E. Lynch, X. Wang, and C. M. Hoffmann, J. Appl. Crystallogr. \textbf{49}, 497 (2016).

\bibitem{Willis2000} A. Wills, Physica B \textbf{276–277}, 680 (2000).

\bibitem{Carvajal1993} J. Rodríguez-Carvajal, Physica B \textbf{192}, 55 (1993).

\bibitem{land} Landwehr G. and Rashba, E. I., \textit{Landau Level Spectroscopy: Modern Problems in Condensed Matter Sciences} Vol. 27.2 (North-Holland, (1991).

\bibitem{shoe} Shoenberg, D., \textit{Magnetic Oscillations in Metals.} (Cambridge Univ. Press, 1984).

\bibitem{fermisurfer} Kawamura, M., Comput. Phys. Commun. \textbf{239}, 197 (2019).

\bibitem{Shuye2013} Shyue Ping Ong, William Davidson Richards, Anubhav Jain, Geoffroy Hautier, Michael Kocher, Shreyas Cholia, Dan Gunter, Vincent Chevrier, Kristin A. Persson, Gerbrand Ceder. Python Materials Genomics (pymatgen) : A Robust, Open-Source Python Library for Materials Analysis. Computational Materials Science, \textbf{68}, 314–319, (2013).

\end{thebibliography}

\begin{thebibliography}{99}

\bibitem{Wang2018} Y.-Y. Wang, Yi-Yan Wang, Sheng Xu, Lin-Lin Sun, and Tian-Long Xia, Quantum oscillations and coherent interlayer transport in a new topological Dirac semimetal candidate YbMnSb$_2$, Phys. Rev. Materials, \textbf{2}, 021201(R) (2018). 

\bibitem{Clark2916} T. M. Michels-Clark, A. T. Savici, V. E. Lynch, X. Wang, and C. M. Hoffmann, J. Appl. Crystallogr. \textbf{49}, 497 (2016).
\bibitem{Willis2000} A. Wills, Physica B \textbf{276–277}, 680 (2000).
\bibitem{Carvajal1993} J. Rodríguez-Carvajal, Physica B \textbf{192}, 55 (1993).
\bibitem{Kealhofer2018}	R. Kealhofer, S. Jang, S. M. Griffin, C. John, K. A. Benavides, S. Doyle, T. Helm, P. J. W. Moll, J. B. Neaton, J. Y. Chan, J. D. Denlinger, and J. G. Analytis, Observation of a two-dimensional Fermi surface and Dirac dispersion in YbMnSb$_2$, Phys. Rev. B. \textbf{97}, 045109 (2018). 

\bibitem{Soh2021} J.-R. Soh, S. M. Tobin, Hao Su, I. Zivkovic, B. Ouladdiaf, A. Stunault, J. Alberto Rodríguez-Velamazán, K. Beauvois, Y. Guo, and A. T. Boothroyd, Magnetic structure of the topological semimetal YbMnSb$_2$, Phys. Rev. B, \textbf{104}, L161103 (2021).

\bibitem{Johnston} David C. Johnston, The puzzle of high temperature superconductivity in layerediron pnictides and chalcogenides, Advances in Physics, \textbf{59}, 803 (2010)

\bibitem{Kresse1996} G. Kresse and J. Furthm\"uller, Phys. Rev. B \textbf{54}, 11169 (1996).

\bibitem{Perdew1996} J. P. Perdew, K. Burke, and M. Ernzerhof, Phys. Rev. Lett. \textbf{77}, 3865 (1996).

\bibitem{Bloechl1994} P. E. Bl\"ochl, Phys.Rev.B \textbf{50}, 17953 (1994).

\bibitem{Dudarev1998} S. L. Dudarev, G. A. Botton, S. Y. Savrasov, C. J. Humphreys, and A. P. Sutton, Phys. Rev. B \textbf{57}, 1505 (1998).

\end{thebibliography}
\end{document}